\documentclass[final,hidelinks]{siamart190516}

\usepackage{amssymb,amsfonts,amsmath}

\usepackage{algorithm,algorithmicx,algpseudocode}

\usepackage{blkarray}

\usepackage[section]{placeins}

\usepackage{xparse,xcolor,cleveref,tikz}
\usetikzlibrary{positioning}
\Crefname{ALC@unique}{Line}{Lines}

\definecolor{darkgreen}{RGB}{0,128,0}
\definecolor{darkblue}{RGB}{0,0,128}
\definecolor{darkred}{RGB}{128,0,0}

\NewDocumentCommand \prArg{mm}
{\left(
\IfNoValueTF{#2}{#1}{#1\,\middle|\,#2}
\right)}

\NewDocumentCommand \newProbabilityFormat{r<>m}
{
	\NewDocumentCommand #1 {e{_}e{^}>{\SplitArgument{1}{|}}r()}
	{
		\IfNoValueTF{##1}
		{
			\IfNoValueTF{##2}
			{#2\prArg##3}
			{#2^{##2}\prArg##3}
		}
		{
			\IfNoValueTF{##2}
			{#2_{##1}\prArg##3}
			{#2_{##1}^{##2}\prArg##3}
		}
	}
}

\NewDocumentCommand \fFunction {m} {{#1}}
\NewDocumentCommand \fSet {m} {\mathcal{#1}}

\NewDocumentCommand \newScalar{r<>m}
{
	\NewDocumentCommand #1 {} {{#2}}
}

\NewDocumentCommand \newVector{r<>m}
{
	\NewDocumentCommand #1 {} {\boldsymbol{#2}}
}

\NewDocumentCommand \newMatrix{r<>m}
{
	\NewDocumentCommand #1 {} {\boldsymbol{#2}}
}

\NewDocumentCommand \newProbability{r<>m}
{
	\newProbabilityFormat<#1>{#2}
}

\NewDocumentCommand \newFunction{r<>m}
{
	\NewDocumentCommand #1 {} {\fFunction{#2}}
}

\NewDocumentCommand \newSet{r<>m}
{
	\NewDocumentCommand #1 {} {\fSet{#2}}
}

\NewDocumentCommand \newAlgorithmTitle{r<>m}
{
	\NewDocumentCommand #1 {} {\texttt{#2} }
}

\makeatletter
\newcommand{\thickhline}{%
    \noalign {\ifnum 0=`}\fi \hrule height 1pt
    \futurelet \reserved@a \@xhline
}
\makeatother

\DeclareMathOperator{\expect}{\mathbb{E}}
\DeclareMathOperator{\var}{\mathrm{Var}}

\NewDocumentCommand \norm {m}{	\left\lVert#1\right\rVert}

\NewDocumentCommand \reals {} {\mathbb{R}}

\newVector<\vA>{a}
\newVector<\vB>{b}
\newVector<\vC>{c}
\newVector<\vD>{d}
\newVector<\vE>{e}
\newVector<\vF>{f}
\newVector<\vG>{g}
\newVector<\vH>{h}
\newVector<\vI>{i}
\newVector<\vJ>{j}
\newVector<\vK>{k}
\newVector<\vL>{l}
\newVector<\vM>{m}
\newVector<\vN>{n}
\newVector<\vP>{p}
\newVector<\vQ>{q}
\newVector<\vR>{r}
\newVector<\vS>{s}
\newVector<\vT>{t}
\newVector<\vU>{u}
\newVector<\vV>{v}
\newVector<\vW>{w}
\newVector<\vX>{x}
\newVector<\vY>{y}
\newVector<\vZ>{z}
\newVector<\vAlpha>{\alpha}
\newVector<\vBeta>{\beta}
\newVector<\vGamma>{\gamma}
\newVector<\vDelta>{\delta}
\newVector<\vEpsilon>{\varepsilon}
\newVector<\vZeta>{\zeta}
\newVector<\vEta>{\eta}
\newVector<\vTheta>{\theta}
\newVector<\vKappa>{\kappa}
\newVector<\vLambda>{\lambda}
\newVector<\vMu>{\mu}
\newVector<\vNu>{\nu}
\newVector<\vXi>{\xi}
\newVector<\vPi>{\pi}
\newVector<\vRho>{\rho}
\newVector<\vSigma>{\sigma}
\newVector<\vTau>{\tau}
\newVector<\vUpsilon>{\upsilon}
\newVector<\vPhi>{\varphi}
\newVector<\vChi>{\chi}
\newVector<\vPsi>{\psi}
\newVector<\vOmega>{\omega}

\newMatrix<\mA>{A}
\newMatrix<\mB>{B}
\newMatrix<\mC>{C}
\newMatrix<\mD>{D}
\newMatrix<\mE>{E}
\newMatrix<\mF>{F}
\newMatrix<\mG>{G}
\newMatrix<\mH>{H}
\newMatrix<\mI>{I}
\newMatrix<\mJ>{J}
\newMatrix<\mK>{K}
\newMatrix<\mL>{L}
\newMatrix<\mM>{M}
\newMatrix<\mN>{N}
\newMatrix<\mP>{P}
\newMatrix<\mQ>{Q}
\newMatrix<\mR>{R}
\newMatrix<\mS>{S}
\newMatrix<\mT>{T}
\newMatrix<\mU>{U}
\newMatrix<\mV>{V}
\newMatrix<\mW>{W}
\newMatrix<\mX>{X}
\newMatrix<\mY>{Y}
\newMatrix<\mZ>{Z}
\newMatrix<\mGamma>{\Gamma}
\newMatrix<\mDelta>{\Delta}
\newMatrix<\mTheta>{\Theta}
\newMatrix<\mLambda>{\Lambda}
\newMatrix<\mXi>{\Xi}
\newMatrix<\mPi>{\Pi}
\newMatrix<\mSigma>{\Sigma}
\newMatrix<\mUpsilon>{\Upsilon}
\newMatrix<\mPhi>{\Phi}
\newMatrix<\mPsi>{\Psi}
\newMatrix<\mOmega>{\Omega}

\renewcommand{\thepage}{C\arabic{page}}

\title{
Randomized Projection for Rank-Revealing Matrix Factorizations and Low-Rank Approximations
\thanks{
Revised from Randomized QR with Column Pivoting for submission to SIGEST
\funding{The original work was support by NSF award 1319312. This revised version was supported by NSF award 1760316 and 
by the Laboratory Directed Research and Development program at Sandia National Laboratories.} 
}
}

\author{Jed A. Duersch\thanks{Sandia National Laboratories, Livermore, CA 94550, United States 
(\href{mailto:jaduers@sandia.gov}{jaduers@\allowbreak sandia.gov})}
\and Ming Gu
\thanks{Department of Mathematics, University of California, Berkeley, CA 94720 (\email{mgu@berkeley.edu}).}
}

\begin{document}
\maketitle

\begin{abstract}
Rank-revealing matrix decompositions provide an essential tool in spectral analysis of matrices, including the Singular Value Decomposition (SVD) and related low-rank approximation techniques.
QR with Column Pivoting (QRCP) is usually suitable for these purposes, but it can be much slower than the unpivoted QR algorithm.
For large matrices, the difference in performance is due to increased communication between the processor and slow memory, which QRCP needs in order to choose pivots during decomposition.
Our main algorithm, Randomized QR with Column Pivoting (RQRCP),
uses randomized projection to make pivot decisions from a much smaller sample matrix, which we can construct to reside in a faster level of memory than the original matrix.
This technique may be understood as trading vastly reduced communication for a controlled increase in uncertainty during the decision process.
For rank-revealing purposes, the selection mechanism in RQRCP produces results that are the same quality as the standard algorithm,
but with performance near that of unpivoted QR (often an order of magnitude faster for large matrices).
We also propose two formulas that facilitate further performance improvements.
The first efficiently updates sample matrices to avoid computing new randomized projections.
The second avoids large trailing updates during the decomposition in truncated low-rank approximations.
Our truncated version of RQRCP also provides a key initial step in our truncated SVD approximation, TUXV.
These advances open up a new performance domain for large matrix factorizations that will support efficient problem-solving techniques for challenging applications in science, engineering, and data analysis.
\end{abstract}




\headers{Randomized Projection for Rank-Revealing Factorizations}{Jed A. Duersch and Ming Gu}

\newMatrix<\mQh>{\hat{Q}}
\newMatrix<\mRh>{\hat{R}}
\newMatrix<\mAh>{\hat{A}}
\newVector<\vAh>{\hat{a}}
\newMatrix<\mBh>{\hat{B}}
\newVector<\vBh>{\hat{b}}
\newMatrix<\mOmegah>{\hat{\Omega}}
\newVector<\vOmegah>{\hat{\omega}}

\newAlgorithmTitle<\aRqrcp>{RQRCP}
\newAlgorithmTitle<\aTrqrcp>{TRQRCP}
\newAlgorithmTitle<\aTuxv>{TUXV}
\newAlgorithmTitle<\aRsrqrcp>{RSRQRCP}
\newAlgorithmTitle<\aQrThree>{DGEQRF}
\newAlgorithmTitle<\aQrcpThree>{DGEQP3}
\newAlgorithmTitle<\aQrTwo>{DGEQR2}
\newAlgorithmTitle<\aSvd>{DGESVD}
\newAlgorithmTitle<\aQrcp>{QRCP}
\newAlgorithmTitle<\aQr>{QR}

\NewDocumentCommand \abs {m} {\left| #1 \right|}
\NewDocumentCommand \eps {} {\varepsilon}

\section{Introduction}

QR with Column Pivoting (QRCP) is a fundamental kernel in numerical linear algebra that broadly supports scientific analysis.
As a rank-revealing matrix factorization, QRCP provides the first step in efficient implementations of spectral methods such as the eigenvalue decomposition
and Principal Component Analysis (PCA), also called the Singular Value Decomposition (SVD)~\cite{Higham2000}.
QRCP also plays a key role in least-squares approximation~\cite{Chan1992} and stable basis extraction~\cite{Stathopoulos2002, Hetmaniuk2006, Duersch2018} for other important algorithms.
These methods allow us to form compressed representations of linear operators by truncation while retaining dominant features that facilitate analytic capabilities that would otherwise be impractical for large matrices.
In the field of data science, PCA and its generalizations~\cite{Vidal2005} support unsupervised machine learning techniques to extract salient features in two-dimensional numerical arrays.
Randomized methods have been extended further to support analysis of multidimensional data using tensor decompositions~\cite{Battaglino2018, Hong2020}.

QRCP builds on the QR decomposition, which expresses a matrix $\mA$ as the product of an orthogonal matrix $\mQ$ and a right-triangular factor $\mR$ as $\mA = \mQ \mR$.
Unlike the LU decomposition, or Gaussian elimination, QR always exists and may be stably computed regardless of the conditioning of $\mA$.
Furthermore, finely tuned library implementations use a blocked algorithm that operates with the communication efficiency of matrix-matrix multiply, or level-3 kernels in the Basic Linear Algebra Subprograms (BLAS-3).
The standard QR decomposition is not, however, suitable for rank detection or low-rank approximations.
These applications require a column permutation scheme to process more representative columns of $\mA$ earlier in the decomposition~\cite{Chan1987,Bischof1991a}.
A permutation matrix $\mP$ encodes these pivoting decisions so that the decomposition becomes $\mA \mP = \mQ \mR$.

The basic QRCP approach selects an unfactorized column with a maximal 2-norm of components that do not reside within the span of previously factorized columns.
This heuristic is a greedy algorithm attempting to maximize the sequence of partial determinants in $\mR$,
which is typically adequate for rank detection with a few notable rare exceptions, such as the Kahan matrix~\cite{Golub2013}.
The critical drawback, however, is that these computations suffer a substantial increase in communication between slow (large) and fast (small) levels of memory,
especially for large matrices.
Each pivot decision requires at least one matrix-vector multiply in series, thus limiting overall efficiency to that of level-2 kernels in the Basic Linear Algebra Subprograms (BLAS-2).

Our primary motivation to examine efficient rank-revealing algorithms is that, for large matrices or algorithms that require frequent use of QRCP, the communication bottleneck becomes prohibitively expensive
and impedes utilization of this important set of analytic tools.

\subsection{Our Contributions}

Randomized projection allows us to reduce communication complexity at the cost of increasing uncertainty regarding latent 2-norms (and inner products) among columns in a large matrix.
Randomized QR with Column Pivoting (RQRCP), shown in \cref{alg:RQRCP}, harnesses this trade-off to obtain full blocks of pivot decisions that can be applied to $\mA$ all at once.
We do this by drawing a Gaussian Independent Identically Distributed (GIID) matrix, $\mOmega$, and compressing columns of $\mA$ into a sample matrix, $\mB = \mOmega \mA$.
The sample matrix retains enough information about the original matrix for us to obtain a full block of pivoting decisions, while requiring far less communication to do so.
Because we construct the sample to contain far fewer rows than the original, it resides in a faster level of memory than the full matrix.
Having a full block of pivots allows us to update the matrix that becomes $\mR$ with blocked BLAS-3 operations in the same fashion as unpivoted QR,
thus entirely eliminating the communication bottleneck of BLAS-2 operations encountered in the standard algorithm.

We show how rank-revealing decompositions are well suited to this approach because each pivot decision must merely avoid 
selecting a column containing a relatively small component orthogonal to the span of previous pivots.
As many columns are often suitable for this purpose, the use of precise 2-norm computations in standard QRCP is unnecessarily.
Our approach has been adopted in subsequent work in computing matrix factorization-based low-rank approximations on sequential and parallel platforms~\cite{EVBK2019,Martinsson2019,Martinsson2017a,Martinsson2020,Xiao2017}. 

We also propose a sample update formula that reduces the number of BLAS-3 operations required to process a full matrix.
Given a block of pivots, updating $\mR$ with blocked Householder reflections requires two matrix-matrix multiplications.
Without a formula to update the sample, we would need a new sample matrix after each block and one additional matrix-matrix multiply.
Instead, we utilize computations that are needed to update $\mR$ to also update $\mB$ into a suitable sample for the next decision block.

RQRCP naturally extends to a truncated formulation, described in \cref{alg:TRQRCP}, that further reduces communication by avoiding trailing updates.
For low-rank matrix approximations, this algorithm requires only one block matrix-matrix multiply per update.
We accomplish this by storing reflector information in the compact WY notation~\cite{Puglisi1992}, which allows us to construct each block of $\mR$ without intermediate updates.
Moreover, this algorithm serves as a key initial step in our approximation of the truncated SVD described in \cref{alg:TUXV}, which is a variant of Stewart's QLP algorithm~\cite{Stewart1999}.

\Cref{sec:related} will discuss the nature of the communication bottleneck and related approaches to address it.
\Cref{sec:sampling} will analyze sample-based pivoting as the maximization of expected utility and derive our main result, RQRCP.
\Cref{sec:truncated} will derive and explain our truncated algorithms for low-rank approximation.
\Cref{sec:experiments} will provide numerical experiments that explore the performance and decomposition quality of these approaches.
\Cref{sec:conclusion} will offer concluding remarks.

\section{Related Work}
\label{sec:related}

In order to understand how our algorithms improve performance, we first review the reasons why additional communication could not be avoided with previous approaches.
Given a large matrix $\mA \in \reals^{m \times n}$, the QRCP decomposition can be computed using \cref{alg:QRCP}.
This algorithm is composed of a sequence of Householder reflections.
To review, a reflector $\vY$ is formed from a particular column, say $\vA$, by subtracting the desired reflection (which must have the same 2-norm) from the current form, such as $\vY = \vA - \left( -\text{sign}(\vA_1) \norm{\vA}_2 \vE_1 \right)$.
The negative sign of the leading element of $\vA$ is used to ensure that the reflector satisfies $\norm{\vY}_2 \geq \norm{\vA}_2$ for numerical stability.
The corresponding reflection coefficient is $\tau = 2 / \vY^T \vY$, which yields the Householder reflection $\mH = \mI - \vY \tau \vY^T$.

In the algorithms that follow, an intermediate state of an array or operator at the end of iteration $j$ is denoted by superscript $(j)$ to emphasize when updates overwrite a previous state.
In contrast, an element computed on iteration $j$ that remains accessible in some form is denoted with a simple subscript.

\begin{algorithm}[h]
\caption{QRCP with BLAS-2 Householder reflections.}
\label{alg:QRCP}
\begin{algorithmic}[1]
\Require \Statex $\mA$ is $m \times n$.
\Ensure
\Statex $\mQ$ is an $m \times m$ orthogonal matrix.
\Statex $\mR$ is an $m \times n$ right triangular matrix, magnitude of diagonals nonincreasing.
\Statex $\mP$ is an $n \times n$ permutation matrix such that $\mA \mP = \mQ \mR$.
\Function {$[\mQ,\mR,\mP]$}{QRCP}{$\mA$}
\State Compute initial column 2-norms which become trailing column norms.
\For{$j=1,2,\ldots, k$, where $k=\min(m,n)$}
   \State Find index $p_j$ of the column with maximum trailing 2-norm.
   \State Swap columns $j$ and $p_j$ with permutation $\mP_j$.
   \State Form Householder reflection $\mH_j = \mI - \vY_j \tau_j \vY_j^T$ from new column.
   \State\label{line:qrcp_reflect}  Apply reflection $\mA^{(j)} = \mH_j (\mA^{(j-1)} \mP_j)$.
   \State\label{line:qrcp_update} Update trailing column norms by removing the contribution of row $j$.
\EndFor
\State $\mQ = \mH_1 \mH_2 \ldots \mH_k$ is the product of all reflections.  \State $\mR = \mA^{(k)}$.
\State $\mP = \mP_1 \mP_2 \ldots \mP_k$ is the aggregate column permutation.
\EndFunction
\end{algorithmic}
\end{algorithm}

At the end of iteration $j$ we can represent the matrix $\mA$ as a partial factorization using the permutation $\mP^{(j)} = \mP_1 \ldots \mP_j$, the composition of column swaps so far, and the analogous composition of Householder reflections, $\mQ^{(j)} = \mH_1 \ldots \mH_j$, to obtain 
\[
        \mA \mP^{(j)} = \mQ^{(j)}
                \left[ \begin{array}{cc}
                         \mR^{(j)}_{11} & \mR^{(j)}_{12} \\
                         0      & \mAh^{(j)} \\
                \end{array} \right].
\]
The leading $j-1$ entries of the vector $\vY_j$ in line $6$ of Algorithm \ref{alg:QRCP} are $0$. 
The upper-left submatrix $\mR^{(j)}_{11} \in \reals^{j \times j}$ is right-triangular.
Likewise, $\mR^{(j)}_{12} \in \reals^{j \times j}$ completes the leading $j$ rows of $\mR$.
It only remains to process the trailing matrix $\mAh^{(j)}$.
On the next iteration, the 2-norm of the selected column within $\mAh^{(j)}$ becomes the magnitude of the next diagonal element in $\mR^{(j+1)}_{11}$.
We may understand QRCP as a greedy procedure intended to maximize the magnitude of the determinant of $\mR^{(j+1)}_{11}$.
The new determinant magnitude is $| \det \mR^{(j+1)}_{11} | = | \det \mR^{(j)}_{11} | \norm{\mAh^{(j)}(:,p_{j+1})}_2$,
so this scheme has selected the pivot that multiplies the previous determinant by the largest factor.
Note that true determinant maximization would require exchanging prior columns and adjusting the factorization accordingly~\cite{Gu1996}.

Early implementations of QR also relied on BLAS-2 kernels and, thus, gave similar performance results until reflector blocking was employed in QR~\cite{Bischof1987,Schreiber1989}.
To process $b$ columns of an $m \times n$ matrix with BLAS-2 kernels, $O(b m n)$ elements must pass from slow to fast memory.
Blocking improves performance by reducing communication between these layers of memory using matrix-matrix multiply.
Instead of updating the entire matrix with each Householder reflection, transformations are collected into a matrix representation that can be applied using two BLAS-3 matrix-matrix multiplies,
which reduces communication complexity to $O(b m n / M^{3/2})$, where $M$ is the size of fast memory.

In order to produce a correct pivot decision at iteration $j+1$ in QRCP, however,
trailing column norms must be updated to remove the contribution of row $j$, which depends on the Householder transformation $\mH_j$.
At first glance, this update appears to require two BLAS-2 operations on the trailing matrix per iteration.
The first operation computes scaled inner products $\vW_j^T = \tau_j \vY_j^T \mA^{(j-1)} \mP_j $ and the second operation modifies the trailing matrix with the rank-1 update $\mA^{(j)} = \mA^{(j-1)} \mP_j - \vY_j \vW_j^T$.
These two operations cause the communication bottleneck in QRCP.

\subsection{Attempts to Achieve BLAS-3 Performance}
\label{sub:attempts}

Quintana-Ort\'{\i}, Sun, and Bischof~\cite{Quintana1998} were able to halve BLAS-2 operations with the insight that the trailing norm update does not require completing the full rank-1 update on each iteration.
Instead, reflections can be gathered into blocks, as in QR.
This method appears in \cref{alg:QRCP3}.

At the end of iteration $j$, the algorithm has collected a block of reflectors $\mY^{(j)}$.
Reflector $\vY_i$, for $i \leq j$, appears in column $i$ from the diagonal down.
This forms a block reflection $\mQ^{(j)} = \mI - \mY^{(j)} \mT^{(j)} \mY^{(j)T}$, where $\mT^{(j)}$ is an upper triangular $j \times j$ connection matrix that can be solved from $\mY^{(j)}$ so that $\mQ^{(j)}$ is orthogonal.
This algorithm must also collect each corresponding scaled inner product, $\vW_i^T$, which appears as row $i$ in a matrix $\mW^{(j)T}$.
In the compact WY notation, this block of inner products is $\mW^{(j)T} = \mT^{(j)T} \mY^{(j)T} \mA \mP^{(j)}$,
which provides enough information to update row $j$ alone and adjust trailing column norms to prepare for the next pivot selection
\[
    \mA^{(j)}(j, :) = \mA^{(j-1)}(j, :) \mP_j - \mY^{(j)}(j,:) \mW^{(j)T}.
\]

Note, however, that this construction complicates reflector formation.
As before, the next pivot index $p_{j+1}$ is selected and swapped into column $j+1$.
Let this new column be $\vA_{j+1}$.
From rows $j+1$ down, elements of $\vA_{j+1}$ have not been updated with the current block of reflectors.
Before we can form $\vY_{j+1}$, prior transformations must be applied to these rows from the formula $\vAh_{j+1} = \vA_{j+1} - \mY^{(j)} \mW^{(j)T}(:, p_{j+1})$.
An additional step is also required to compute reflector inner products because they must account for reflections that have not been applied to the trailing matrix.
The adjusted formula for these inner products is
\[
    \vW_{j+1}^T = \tau_{j+1} \left( \vY_{j+1}^T \mA^{(j)} - (\vY_{j+1}^T \mY^{(j)}) \mW^{(j)T} \right) \mP_{j+1}.
\]
The reflector and inner product blocks are then updated:
\[
    \mY^{(j+1)} =
    \begin{bmatrix}
        \mY^{(j)} & \vY_{j+1} \\
    \end{bmatrix}
    \quad\text{and}\quad
    \mW^{(j+1)T} = 
    \begin{bmatrix}
        \mW^{(j)T} \mP_{j+1} \\
        \vW_{j+1}^T \\
    \end{bmatrix}
    \text{.}
\]
Unfortunately, the remaining BLAS-2 operations $\vY_{j+1}^T \mA^{(j)}$ and $\vY_{j+1}^T \mY^{(j)}$ in the inner product computation still dominate slow communication complexity for large matrices.
The entire trailing matrix must pass from slow to fast memory once per iteration.
Consequently, even heavily optimized implementations of blocked QRCP still run substantially slower than blocked QR on both sequential and parallel architectures.

\begin{algorithm}[h]
\caption{QRCP with BLAS-3 reflection blocking.}
\label{alg:QRCP3}
\begin{algorithmic}[1]
\Require
\Statex $\mA$ is $m \times n$.
\Ensure
\Statex $\mQ$ is an $m \times m$ orthogonal matrix.
\Statex $\mR$ is an $m \times n$ right triangular matrix, diagonals in nonincreasing magnitude order.
\Statex $\mP$ is an $n \times n$ permutation matrix such that $\mA \mP=\mQ \mR$.
\Function {$[\mQ,\mR,\mP]=$ QRCP}{$\mA$}
\State Compute initial column 2-norms, which will become trailing column norms.
\For{$i=0,b,2b \ldots, $ where $b$ is block size.}
\For{$j=i+1,i+2,\ldots \min(i+b,k)$, where $k=\min(m,n)$.}
   \State Find index $p_j$ of the column with maximum trailing 2-norm.
   \State Apply permutation $\mP_j$ swapping column $j$ with $p_j$.
   \State \textbf{Update column $j$ with prior reflections in this block.}
   \State Form reflector $\vY_j$ and $\tau_j$ from new column $j$.
   \State \textbf{Compute adjusted reflector inner products $\vW_j^T$.}
   \State \textbf{Update row $j$ with all reflections in this block.}
   \State Update trailing column norms by removing the contribution of row $j$.
\EndFor
\State \textbf{Apply block reflection to trailing matrix.}
\EndFor
\State $\mQ = \mI - \mY_k \mT_k \mY_k^T$ where $\mT_k$ can be recovered from $\mY_k$ and $\tau_1, \ldots, \tau_k$.
\State $\mR = \mA^{(k)}$.
\State $\mP = \mP_1 \mP_2 \ldots \mP_k$ is the aggregate column permutation.
\EndFunction
\end{algorithmic}
\end{algorithm}

\subsection{Communication Avoiding Rank-Revealing QR}

Several mechanisms have been put forward to avoid repeating full passes over the trailing matrix on each iteration.
Bischof~\cite{Bischof1991} proposed pivoting restricted to local blocks and Demmel, et al.~\cite{Demmel2012,Demmel2015} propose a procedure called Communication Avoiding Rank-Revealing QR (CARRQR).
CARRQR proceeds by partitioning the trailing matrix into $\mathcal{P}$ subsets of columns that are processed independently and possibly simultaneously.
From within each column subset, $b$ candidate pivots are selected using QRCP.
Adjacent subsets of candidates are then combined to form $\frac{1}{2}\mathcal{P}$ subsets of $2b$ candidates.
This procedure continues, using QRCP to filter $b$ candidates per subset followed by merging results into $\frac{1}{4}\mathcal{P}$ subsets and so on, until only one subset of $b$ candidates remains.
The trailing matrix is then updated as before, with blocked reflections.

We now examine several practical constraints in implementing CARRQR.
First, the reflectors $\mY$, inner products $\mW^T$, and leading rows of $\mR$ must be stored separately from the original matrix for each independently processed subset of columns.
Furthermore, one must employ a version of QRCP that avoids the trailing update,
because the final reflectors are unknown until the last selection stage.
Any intermediate changes to the original columns would have to be undone before the final transformations can be correctly processed.
In contrast, QRCP can be written to convert columns into reflectors, storing the results in the same array as the input on the strictly lower triangle portion of the matrix.
Likewise, $\mR$ can be stored in place of the input on the upper triangle.

Depending on the initial column partition, CARRQR performs up to 2 times as many inner products as QRCP per block iteration.
Note that as the reflector index $j$ increases, the total number of inner products of the form $\vY_{j+1}^T \vY_{j+1}$, $\vY_{j+1}^T \mY^{(j)}$ and $\vY_{j+1}^T \mA^{(j)}$ remains constant.
Therefore, if the $i$th subset contains $n_i$ columns, $b n_i$ inner products will be required to produce $b$ candidates.
Letting $n_1 + n_2 + \cdots + n_\mathcal{P} = n$ on the first stage of refinement, summing over all of the subsets gives $b n$ inner products to produce $\mathcal{P}$ subsets of $b$ candidates.
QRCP requires the same computational complexity to produce $b$ final pivots.
Assuming that the number of candidates is at least halved for each subsequent stage of refinement in CARRQR, it easily follows that no more than $2 b n$ inner products will be computed in total.

Despite increased computational complexity, CARRQR is intended to benefit from better memory utilization and better parallel scalability.
If each column subset is thin enough to fit in fast memory, then slow communication is eliminated between iterations of $j$.
The only remaining slow communication transmits pivot candidates between stages of refinement.

Unfortunately, writing and tuning CARRQR is nontrivial.
We implemented this algorithm and found that it ran slightly slower than the LAPACK implementation of \cref{alg:QRCP3}, called DGEQP3, on a shared-memory parallel machine.
We believe this was mainly due to inefficient parallelization in the final stages of refinement.
We assigned each column subset to a different processor, which then worked independently to produce candidates.
This approach was attractive because it did not require communication between processors during each filtration stage.
However, despite communication efficiency, this technique can only engage as many processors as there are column subsets.
Most processors are left idle during the final stages of refinement.
A second problem with this approach occurres when the matrix is too tall.
In such cases, it is not possible to select column subsets that are thin enough to fit into fast memory.
An efficient implementation would need alternative or additional workload-splitting tactics to use all processors at every stage of refinement.

As we will discuss in the next section, the method we propose also gathers pivots into blocks, which are then applied to the trailing matrix.
Our method, however, improves performance by reducing both the communication and computational complexity needed to form a block of pivots.

\section{Randomized Projection for Sample Pivoting}
\label{sec:sampling}

Sampling via randomized projection has proven to be beneficial for a variety of applications in numerical linear algebra~\cite{Liberty2007,Woolfe2008,Rokhlin2010,Halko2011,Mahoney2011,Martinsson2011}.
Sampling reduces communication complexity via dimensional reduction, while simultaneously maintaining a safe degree of uncertainty in the approximations that follow.
This technique is typically framed using the Johnson--Lindenstrauss lemma~\cite{Johnson1984}.
Let $\vA_j$ represent the $j$th column of $\mA$ for $j=1, 2, \ldots, n$.
There exists a distribution over $\ell \times m$ matrices such that a randomly drawn matrix $\mOmega$ yields a lower dimensional embedding $\vB_j = \mOmega \vA_j$,
with high probability of preserving distances within a relative error $\eps$.
Because we would like to use the sample to obtain a block of $b$ pivots while controlling the degree of uncertainty in the true norms of $\mA$,
we include padding $p$ in the sample rank so that $\ell = b + p$.
When $\mOmega$ is GIID, the expected 2-norms and variance are
\[
    \expect_{\mOmega}\left[\norm{\vB_j}_2^2\right] = \ell \norm{\vA_j}_2^2
    \quad\text{and}\quad
    \var_{\mOmega}\left[\norm{\vB_j}_2^2\right] = 2 \ell \norm{\vA_j}_2^4.
\]
If we let $\vA_0=0$ and $\vB_0=0$, then we can express the probability of satisfying the relative error bounds as
\[
P\left( \abs{\frac{\norm{\vB_j-\vB_i}_2^2}{\ell \norm{\vA_j-\vA_i}_2^2} - 1} \leq \eps \right) \geq 1 - 2 \exp\left(\frac{-\ell \eps^2}{4}(1 - \eps)\right),
\]
where $0 < \eps < \frac{1}{2}$ and $i,j = 0,1, \ldots, n$.
Capturing the norm of the difference between columns also implies coherence among inner product approximations
by taking $(\vB_1 - \vB_2)^T (\vB_1 - \vB_2)  = \norm{\vB_1}_2^2 - 2 \vB_1^T \vB_2 + \norm{\vB_2}_2^2$.
Thus, each component of $\mB$ within a subspace defined by a few of its columns approximates the corresponding component of $\mA$.

\subsection{Bayesian Analysis}
Bayesian inference is not often used in numerical linear algebra.
In this case, it allows us to obtain an exact expression for the uncertainty in column norms,
which rigorously frames the amount of padding $p$ that is needed to satisfy a relative error bound with a desired probability of success.

We begin by expressing a single column $\vA$ as its 2-norm multiplied by a unit vector $\vQ$, so that $\vA = \norm{\vA}_2 \vQ$.
Because a GIID matrix is invariant in distribution under independent orthogonal transformations,
we can construct an orthogonal matrix $\mQ = [\vQ\; \mQ_{\bot}]$ and write $\mOmega$ as 
\[
    \mOmega = \left[ \begin{array}{cc}
                \vOmegah & \mOmegah \\
                \end{array} \right]
                \left[ \begin{array}{c}
                        \vQ^T \\
                        \mQ_{\bot}^T \\
                \end{array} \right],
\]
where both the leading column $\vOmegah$ and remaining columns $\mOmegah$ are GIID.
It easily follows that each element of the sample column, $\vB = \norm{\vA}_2 \vOmegah$, is normally distributed with mean $0$ and latent variance $\norm{\vA}_2^2$.

Inferring variance from a normal distribution with a known mean is a standard problem in Bayesian statistics.
The likelihood probability distribution is written $p(\vB \mid \norm{\vA}_2^2) \equiv \mathcal{N}(\vB \mid 0, \norm{\vA}_2^2 \mI)$ where $\mI$ is the $\ell \times \ell$ identity.
Jeffreys proposed the maximally uninformative prior for an unknown variance, $p(\norm{\vA}_2^2) \equiv \norm{\vA}_2^{-2}$, which is invariant under scaling and power transformations~\cite{Kass1996}.
We apply Bayes' theorem, $p(\norm{\vA}_2^2 \mid \vB) \propto p(\vB \mid \norm{\vA}_2^2) p(\norm{\vA}_2^2)$, to obtain the posterior distribution.
Normalization results in the inverse gamma distribution
\[
p(\norm{\vA}_2^2 \mid \norm{\vB}_2^2) \equiv \frac{\left(\frac{\norm{\vB}_2^2}{2}\right)^{\ell/2}}{\Gamma(\frac{\ell}{2}) \norm{\vA}_2^{-\ell-2}} \exp\left( \frac{-\norm{\vB}_2^2}{2 \norm{\vA}_2^2} \right).
\]
Consequently, we can cast each pivoting decision as the maximizer of expected utility, where utility is taken to be the latent 2-norm squared
\[
    \expect_{p(\norm{\vA}_2^2 \mid \norm{\vB}_2^2) }\left[\norm{\vA}_2^2\right] = \frac{\norm{\vB}_2^2}{\ell - 2}.
\]
Since expected utility is monotonic in the sample column norms, we simply choose the maximum as in QRCP.

More importantly, the posterior distribution clearly relates sample rank $\ell$ to uncertainty.
In order to capture the probability distribution of the latent relative error,
we can change variables to express the latent column norm as a fraction $\phi$ of the expectation, $\norm{\vA}_2^2 = \frac{\phi}{\ell - 2} \norm{\vB}_2^2$.
This gives analytic expressions for both the probability distribution function and the cumulative distribution function of the relative scaling
\[
p(\phi \mid \ell) \equiv \frac{ \left( \frac{\ell-2}{2} \right)^{\ell/2} }{ \Gamma(\frac{\ell}{2}) \phi^{\ell/2+1}} \exp\left( \frac{-(\ell - 2)}{2 \phi} \right)
\quad\text{and}\quad
P(\phi < \tau) = \frac{\Gamma(\frac{\ell}{2},\frac{\ell - 2}{2 \tau} )}{ \Gamma(\frac{\ell}{2})},
\]
respectively.
The numerator of the cumulative distribution is the upper incomplete gamma function and normalization results in the regularized gamma function.

\begin{figure}[h]
\centering
\includegraphics[width=0.69\textwidth]{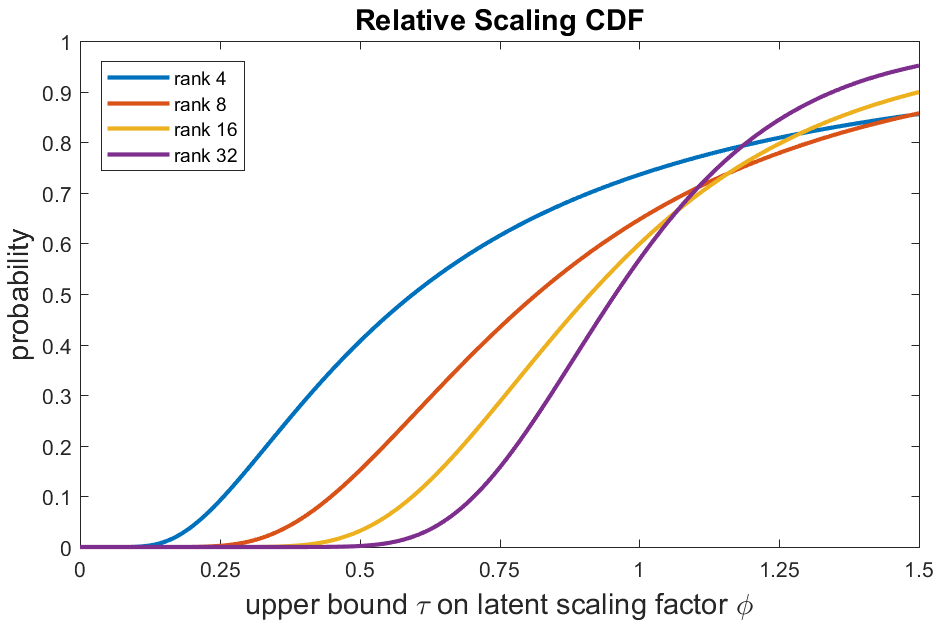}
\caption{Cumulative distribution function for latent relative scaling factor.
For sample rank 4, the probability that the latent 2-norm squared is less than $1/8$ of the expectation is $0.30\%$.
For sample rank 8, the probability that it is less than $1/4$ is $0.23\%$.
For sample rank 32, the probability that it is less than $1/2$ is $0.20\%$.
}
\label{fig:latentCdf}
\end{figure}

Rank-revealing decompositions must avoid selecting columns that are already well approximated by components in the span of previous pivots,
i.e columns with small trailing norms.
As such, we only care about the probability that a sample column radically overestimates the true column norm.
The CDF plotted in \cref{fig:latentCdf} shows the probability that the relative scaling $\phi$ falls below a specified upper bound $\tau$ for several choices of $\ell$.
Our experiments provide good results using padding $p=8$ with a block size $b=32$ so that the effective sample rank satisfies $8 < \ell \leq 40$ for each pivot decision.
Note that this analysis also holds for linear combinations of columns in $\mA$ and $\mB$, provided that those linear combinations are independent of $\mB$.
An in-depth reliability and probability analysis has appeared in Xiao, Gu, and Langou~\cite{Xiao2017}.
In particular, they show that in the case of decaying singular values in the matrix $\mA$, a nearly optimal low-rank approximation can be computed with the QR factorization with a slight modification in the strategy used to choose $\mP$. 
 
\subsection{Sample QRCP Distribution Updates}
\label{subsec:SSRQRCP}

We now show how the sample matrix $\mB=\mOmega \mA$ can be used to select a full block of $b$ pivots.
If QRCP is performed on the sample matrix, then at iteration $j$ we can examine $\mB$ as a partial factorization.
We let $\mP^{(j)}$ be the aggregate permutation so far and
represent the accumulated orthogonal transformations applied to $\mB$ as $\mU^{(j)}$, with corresponding intermediate triangular factor $\mS^{(j)}$, as shown below.
We also consider a partial factorization of $\mA$ using the same pivots that were applied to $\mB$.
The corresponding factors of $\mA$ are $\mQ^{(j)}$ and $\mR$.
\[
    \mB \mP^{(j)} = \mU^{(j)} \begin{bmatrix}
        \mS_{11}^{(j)} & \mS_{12}^{(j)} \\
        0            & \mS_{22}^{(j)} \\
    \end{bmatrix}
    \quad\text{ and }\quad
    \mA \mP^{(j)} = \mQ^{(j)} \begin{bmatrix}
        \mR_{11}^{(j)} & \mR_{12}^{(j)} \\
        0            & \mR_{22}^{(j)} \\
    \end{bmatrix}
    \text{.}
\]
Both $\mS_{11}^{(j)}$ and $\mR_{11}^{(j)}$ are upper triangular.
$\mOmega$ can then be expressed as elements $\mOmegah$ in the bases given by $\mU^{(j)}$ and $\mQ^{(j)}$:
\[
    \mOmega = \mU^{(j)} \begin{bmatrix}
        \mOmegah_{11}^{(j)} & \mOmegah_{12}^{(j)} \\
        \mOmegah_{21}^{(j)} & \mOmegah_{22}^{(j)} \\
    \end{bmatrix} \mQ^{(j)T}
    \text{.}
\]
Noting that $\mB \mP^{(j)} = \mOmega \mA \mP^{(j)}$, we have
\begin{equation}
    \begin{bmatrix}
        \mS_{11}^{(j)} & \mS_{12}^{(j)} \\
        0            & \mS_{22}^{(j)} \\
    \end{bmatrix} =
    \begin{bmatrix}
      \mOmegah_{11}^{(j)} \mR_{11}^{(j)} & \mOmegah_{11}^{(j)} \mR_{12}^{(j)}+\mOmegah_{12}^{(j)} \mR_{22}^{(j)} \\
      \mOmegah_{21}^{(j)} \mR_{11}^{(j)} & \mOmegah_{21}^{(j)} \mR_{12}^{(j)}+\mOmegah_{22}^{(j)} \mR_{22}^{(j)} \\
    \end{bmatrix}\text{.}
    \label{eq:sample}
\end{equation}

If $\mS_{11}^{(j)}$ is nonsingular, then both $\mOmegah_{11}^{(j)}$ and $\mR_{11}^{(j)}$ are also nonsingular.
It follows that $\mOmegah_{11}^{(j)} = \mS_{11}^{(j)} \mR_{11}^{(j)-1}$ is upper triangular and $\mOmegah_{21}^{(j)}=0$.
In other words, we have implicitly formed a QR factorization of $\mOmega \mQ^{(j)}$ using the same orthogonal matrix $\mU^{(j)}$.
Finally, the trailing matrix in the sample simplifies to $\mS_{22}^{(j)}=\mOmegah_{22}^{(j)}\mR_{22}^{(j)}$, which is a sample of the trailing matrix $\mR_{22}^{(j)}$ using the compression matrix $\mOmegah_{22}^{(j)}$.
If the permutation $\mP^{(j)}$ were independent of the sample, then $\mQ^{(j)}$ would be formed from $\mA$, independent of $\mOmega$.
Likewise, $\mU^{(j)}$ only depends on the leading $j$ columns of the sample, which are independent of the trailing columns.
Thus $\mOmegah_{22}^{(j)}$ would be GIID.
As such, we may use column norms of $\mS_{22}^{(j)}$ to approximate column norms of $\mR_{22}^{(j)}$ when we select the $(j+1)$st pivot,
so that a full block of pivots can be selected without interleaving any references to $\mA$ or $\mR$ memory.
We note, however, that the permutation may exhibit a subtle dependence on the pivots, which we briefly discuss in \cref{sub:bias}.

Once $b$ pivots have been selected from the sample matrix $\mB$, the corresponding columns of $\mA$ are permuted and processed all at once, as is done in BLAS-3 QR,
thus reducing both the communication and computational complexity associated with selecting a block of $b$ pivots by a factor of $\ell / m$.
More significantly, if the sample matrix $\mB$ fits in fast memory, then slow communication between consecutive pivot decisions is eliminated within each block iteration.
As in BLAS-3 QR, the remaining communication costs are due to the matrix-matrix multiplications needed to perform block reflections.
As such, RQRCP satisfies the BLAS-3 performance standard.

\Cref{alg:SSRQRCP} outlines the full procedure for a sample-based block permutation.
It is acceptable for very low-rank approximations, wherein the required sample is small enough to maintain communication efficiency.
That is, when the desired approximation rank $k$ is small enough to be a single block, $b=k$.
For larger approximations we will resort to a more comprehensive algorithm that includes a sample update formulation that subsumes this version.
Since the single-sample algorithm illuminates the performance advantage gained from this approach, we examine it first.

\begin{algorithm}
\caption{Single-Sample Randomized QRCP.}
\label{alg:SSRQRCP}
\begin{algorithmic}[1]
\Require
\Statex $\mA$ is $m \times n$.
\Statex $k$ is the desired approximation rank.
$k \ll \min(m,n)$.
\Ensure
\Statex $\mQ$ is an $m \times m$ orthogonal matrix in the form of $k$ reflectors.
\Statex $\mR$ is a $k \times n$ truncated upper trapezoidal matrix.
\Statex $\mP$ is an $n \times n$ permutation matrix such that $\mA \mP \approx \mQ(:,1:k) \mR$.
\Function {$[\mQ,\mR,\mP]=$ SingleSampleRQRCP}{$\mA, k$}
\State Set sample rank $l=k+p$ as needed for acceptable uncertainty.

\State Generate random $l \times m$ matrix $\mOmega$.
\State Form the sample $\mB=\mOmega \mA$.
\State \textbf{Get $k$ column pivots from sample $[\cdot,\cdot,\mP]=\aQrcp(\mB)$.}
\State \textbf{Apply permutation $\mA^{(1)}=\mA \mP$.}
\State Construct $k$ reflectors from new leading columns $[\mQ,\mR_{11}]=\aQr(\mA^{(1)}(\texttt{:,1:k}))$.
\State Finish $k$ rows of $\mR$ in remaining columns $\mR_{12}=\mQ(\texttt{:,1:k})^T \mA^{(1)}(\texttt{:,k+1:n})$.
\EndFunction
\end{algorithmic}
\end{algorithm}

\subsection{Sample Bias}
\label{sub:bias}

The bias of an estimator is the difference between the expected value of the estimator and the true value of the quantity being estimated.
In this case, sample column norms are used to estimate true column norms.
As illustrated in the following thought experiment, a subtle form of bias occurs when we use the maximum to select a pivot.

Suppose Jack and Jill roll one six-sided die each.
The expected outcome for each of them is $3.5$.
We then learn that Jill rolled $5$, which was greater than Jack's roll.
This new information reduces Jack's expectation to $2.5$, because it is no longer possible for him to have rolled $5$ or $6$.
Similarly, the act of selecting the largest sample norm creates a dependency with remaining samples by truncating their plausible outcomes.

This effect becomes less pronounced, however, if the decision is more likely to be determined by the true value being estimated rather than a chance sample outcome.
For example, now suppose that Jill rolls three dice and Jack rolls one.
If we only know each person's sum, we can still infer the number of dice each person rolled and select the expected maximum as before.
In $90.7\%$ of trials, Jill's sum will be $7$ or greater, driven by the fact that three dice were rolled.
In most cases, this leaves Jack's potential outcomes unconstrained and his expectation unbiased after we observe the maximum.
Analogously, when a selected column exhibits a norm that is an order of magnitude greater than others,
the bias effect is negligible and we may proceed as though the pivot decision was independent of the sample.

The original version of this paper included analysis of this distribution truncation effect \cite{Duersch2017}.
For our purposes, we proceed as though the progression of sample updates is independent of the pivot decisions,
which is also the approach Xiao, et al.~\cite{Xiao2017} take.
This assumption is potentially problematic when multiple trailing columns have similar norms,
but any such column provides a suitable pivot in that scenario.

\subsection{Blocked Sample Updates}

The sample matrix in \cref{alg:SSRQRCP} is formulated to have rank $\ell = k + p$, where $k$ was both the desired approximation rank and the permutation block size, $b$.
As $k$ increases, however, it becomes inefficient to simply increase the sample rank $\ell$.
In the extreme case, a full decomposition would require a sample just as big as the original matrix.
If we require a decomposition with a larger rank than that which can be efficiently sampled and blocked, that is if the sample rank cannot exceed $\ell = b + p$, but we require $k>b$, then we need to update the sample matrix after each block.
Martinsson~\cite{Martinsson2016} developed an approach in which one simply processes each subsequent block by drawing a new random matrix $\mOmega$ and applying it to each new trailing matrix.
We propose a sample update formulation that does not require multiplying the trailing matrix by a new compression matrix
and reduces BLAS-3 communication in the overall factorization by at least one third.

The update formula we derive is an extension of the implicit update mechanism described in the previous section.
Both \cref{alg:RQRCP} and \cref{alg:TRQRCP}, the truncated variation, will proceed in blocks of pivots.
Bracket superscripts denote the results of a computation that occurred on the indicated block-iteration.
At entry to the first block-iteration, the sample is $\mB^{[0]}=\mOmega^{[0]} \mA^{[0]}$, where $\mA^{[0]}$ is the original matrix.
At the end of block-iteration $J$, the sample will be in the transformed state
\begin{equation}
        \begin{bmatrix}
                \mS_{11}^{[J]} & \mS_{12}^{[J]} \\
    0            & \mS_{22}^{[J]} \\
  \end{bmatrix}
  =
  \begin{bmatrix}
        \mOmegah_{11}^{[J]} & \mOmegah_{12}^{[J]} \\
    0            & \mOmegah_{22}^{[J]} \\
  \end{bmatrix}
  \begin{bmatrix}
        \mR_{11}^{[J]} & \mR_{12}^{[J]} \\
    0            & \mA^{[J]} \\
  \end{bmatrix},
  \label{eq:sampleR}
\end{equation}
just as in \cref{eq:sample}.
$\mS_{11}^{[J]}$ is the leading upper triangle from the partial factorization of the sample and $\mS_{22}^{[J]}$ gives the trailing sample columns.
Likewise $\mR_{11}^{[J]}$ and $\mA^{[J]}$, respectively, give the leading upper triangle and trailing columns that would be obtained by factorizing the original matrix with the same pivots.
By absorbing the transformations $\mU^{[J]T}$ and $\mQ^{[J]}$ into $\mOmega^{[J]}$, we obtained an effective compression matrix $\mOmegah_{22}^{[J]}$,
which had already been implicitly applied to the trailing columns: $\mS_{22}^{[J]} = \mOmegah_{22}^{[J]} \mA^{[J]}$.
The difficulty is $\mS_{22}^{[J]}$ only has rank $p$.
In order to construct a rank $\ell = b + p$ sample of the trailing matrix $\mA^{[J]}$, we need to include $\mOmegah_{12}^{[J]}$ in the updated compression matrix
\[
    \mOmega^{[J]} =
    \begin{bmatrix}
      \mOmegah_{12}^{[J]} \\
      \mOmegah_{22}^{[J]} \\
    \end{bmatrix}
    \quad
    \text{giving}
    \quad
    \mB^{[J]} = \mOmega^{[J]} \mA^{[J]} =
    \begin{bmatrix}
        \mS_{12}^{[J]} - \mOmegah_{11}^{[J]} \mR_{12}^{[J]} \\
      \mS_{22}^{[J]} \\
    \end{bmatrix}.
\]
In other words, the new compression matrix $\mOmega^{[J]}$ is simply $\mU^{[J]T}\mOmega^{[J-1]}\mQ^{[J]}$, with the leading $b$ columns removed.
This new compression matrix does not need to be explicitly formed or applied to the trailing columns $\mA^{[J]}$.
Instead, we form the result implicitly by removing $\mOmegah_{11}^{[J]} \mR_{12}^{[J]}$ from $\mS_{12}^{[J]}$.
Both $\mR_{11}^{[J]}$ and $\mR_{12}^{[J]}$ will be computed in blocked matrix multiply operations using the previous $b$ pivots of $\mA$.
Since $\mOmegah_{11}^{[J]}$ can then be recovered from $\mS_{11}^{[J]}$, we can avoid any direct computations on $\mOmega$.
We only need to update the first $b$ rows of $\mB$, which gives us the sample update formula
\begin{equation}
    \left[ \begin{array}{c}
        \mB_1^{[J]} \\
        \mB_2^{[J]} \\
    \end{array} \right]
    = \left[ \begin{array}{c}
        \mS_{12}^{[J]} - \mS_{11}^{[J]} \mR_{11}^{[J]-1} \mR_{12}^{[J]} \\
        \mS_{22}^{[J]} \\
      \end{array} \right].
      \label{eq:update1}
\end{equation}

Full RQRCP, described in \cref{alg:RQRCP}, can be structured as a modification to blocked BLAS-3 QR.
The algorithm must simply interleave processing blocks of reflectors with permutations obtained from each sample matrix, then update the sample as above.
To obtain a modified version that employs repeated sampling simply replace the sample update with a new sample of the trailing matrix, $\mB^{[J]} = \mOmega^{[J]} \mA^{[J]}$. 

When QRCP is applied to the sample matrix $\mB$, only a partial decomposition is necessary.
The second argument $b$ in the subroutine call $\aQrcp(\mB^{[J]},b)$ indicates that only $b$ column permutations are required.
It is relatively simple to modify the QR algorithm to avoid any unnecessary computation.
After sample pivots have been applied to the array containing both $\mA$ and $\mR$, we perform QR factorization on the new leading $b$ columns of the trailing matrix.
Although this is stated as returning $\mQ^{[J]}$ for convenience, it can be implemented efficiently with the blocked Householder reflections described in \cref{sub:attempts}.
We can then apply reflectors to the trailing matrix and form the sample update $\mB^{[J]}$ to prepare for the next iteration.

\begin{algorithm}[h]
        \caption{Randomized QR with Column Pivoting, RQRCP}
        \label{alg:RQRCP}
        \begin{algorithmic}[1]
                \Require
                        \Statex $\mA$ is $m \times n$.
                    \Statex $k$ is the desired factorization rank.
$k \le \min{(m,n)}$.
                \Ensure
                    \Statex $\mQ$ is an $m \times m$ orthogonal matrix in the form of $k$ reflectors.
                    \Statex $\mR$ is a $k \times n$ upper trapezoidal (or triangular) matrix.
                    \Statex $\mP$ is an $n \times n$ permutation matrix such that $\mA \mP \approx \mQ(\texttt{:,1:k})\mR$.
                \Function {$[\mQ,\mR,\mP]=$ RQRCP}{$\mA,k$}
                \State Set sample rank $\ell = b + p$ as needed for acceptable uncertainty.
                \State Generate random $\ell \times m$ matrix $\mOmega^{[0]}$.
                \State Form the initial sample $\mB^{[0]} = \mOmega^{[0]} \mA^{[0]}$.
                \For{$J=1$, 2, \ldots, $\frac{k}{b}$}
                    \State \textbf{Get $b$ column pivots from sample} $[\mU^{[J]},\mS^{[J]},\mP^{[J]}] = \aQrcp(\mB^{[J-1]},b)$.
                    \State \textbf{Permute $\mA^{[J-1]}$ and completed rows in $\mR$ with $\mP^{[J]}$.}
                    \State Construct $b$ reflectors $[\mQ^{[J]},\mR_{11}^{[J]}] = \aQr(\mA^{[J-1]}(\texttt{:,1:b}))$.
                    \State Finish $b$ rows $\mR_{12}^{[J]}=\mQ^{[J]}(\texttt{:,1:b})^T \mA^{[J-1]}(\texttt{:,b+1:end})$.
                    \State Update the trailing matrix $\mA^{[J]}=\mQ^{[J]}(\texttt{:,b+1:end})^T\!\mA^{[J-1]}(\texttt{:,b+1:end})$.
                    \State \textbf{Update the sample} $\mB_{1}^{[J]}=\mS_{12}^{[J]}-\mS_{11}^{[J]}\mR_{11}^{[J]-1} \mR_{12}^{[J]}$ and $\mB_{2}^{[J]}=\mS_{22}^{[J]}$.
                \EndFor
                \State $\mQ = \mQ^{[1]}  \left[ \begin{array}{cc}
                         \mI_{b}&  \\
                         & \mQ^{[2]} \\
                \end{array} \right]  \ldots  \left[ \begin{array}{cc}
                         \mI_{\left([k/b]-1\right)\,b}&  \\
                         & \mQ^{[k/b]}\\
                \end{array} \right]  $.
                \State $\mP = \mP^{[1]} \left[ \begin{array}{cc}
                         \mI_{b}&  \\
                         & \mP^{[2]} \\
                \end{array} \right]  \ldots  \left[ \begin{array}{cc}
                         \mI_{\left([k/b]-1\right)\,b}&  \\
                         & \mP^{[k/b]}\\
                \end{array} \right]$.
        \EndFunction
        \end{algorithmic}
\end{algorithm}

\section{Truncated Factorizations for Low-Rank Approximations}
\label{sec:truncated}

The trailing matrix is unnecessary for low-rank applications.
We can reformulate RQRCP to avoid the trailing update, rather than computing it and discarding it.
Provided the approximation rank is small ($k\ll\min(m,n)$), the truncated reformulation (TRQRCP) reduces large matrix multiplications by half and completes in roughly half the time.

\subsection{Truncated RQRCP}
Our technique is analogous to the method Quintana-Ort\'{\i}, Sun, and Bischof used to halve BLAS-2 operations in QRCP.
In their version of QRCP, all reflector inner products are computed, but rows and columns are only updated as needed.
In order to compute correct reflector inner products, without having updated the trailing matrix, we need to formulate blocked reflector compositions
\begin{align*}
& (\mI - \mY_1 \mT_1 \mY_1^T)(\mI-\mY_2 \mT_2 \mY_2^T) = \mI - \mY \mT \mY^T, \\
&\text{where}\quad
    \mY =
    \begin{bmatrix}
        \mY_1 & \mY_2 \\
    \end{bmatrix}
    \quad\text{ and }\quad
    \mT =
    \begin{bmatrix}
        \mT_1 & -\mT_1 \mY_1^T \mY_2 \mT_2 \\
        0   & \mT_2 \\
    \end{bmatrix}.
\end{align*}
The corresponding reflector inner products $\mW^T = \mT^T \mY^T \mA$ become
\[
    \mW^T =
    \begin{bmatrix}
        \mW_1^T \\
        \mW_2^T \\
    \end{bmatrix}
    \quad
    \text{with}
    \quad
    \mW_1^T = \mT_1^T \mY_1^T \mA
    \quad
    \text{and}
    \quad
    \mW_2^T = \mT_2^T \left( \mY_2^T \mA - (\mY_2^T \mY_1) \mW_1^T \right)
    \text{.}
\]
If we store these reflector inner products, then we can construct any submatrix of the accumulated transformation $\mAh^{[J]} = \mA - \mY^{[J]} \mW^{[J]T}$ as needed.
Columns that are selected by sample pivots are constructed just before becoming the next reflectors and corresponding rows of $\mR$ are constructed just before being used to update the sample,
as outlined in \cref{alg:TRQRCP}.

\begin{algorithm}[htb]
        \caption{Truncated RQRCP without trailing update}
        \label{alg:TRQRCP}
        \begin{algorithmic}[1]
                \Require
                        \Statex $\mA$ is $m \times n$.
                        \Statex $k$ is the approximation rank.
$k\ll\min(m,n)$.
                \Ensure
                    \Statex $\mQ$ is an $m \times m$ orthogonal matrix in the form of $k$ reflectors.
                    \Statex $\mR$ is a $k \times n$ upper trapezoidal matrix.
                    \Statex $\mP$ is an $n \times n$ permutation matrix such that $\mA \mP \approx \mQ(\texttt{:,1:k}) \mR$.
                \Function {$[\mQ,\mR,\mP]$ TruncatedRQRCP}{$\mA,k$}
                \State Set the sample rank $\ell = b + p$ as needed for acceptable uncertainty.
                \State Generate $\ell \times m$ random matrix $\mOmega^{[0]}$ and sample $\mB^{[0]} = \mOmega^{[0]} \mA{[0]}$.
                \For{$J=1$, 2, \ldots, $\frac{k}{b}$}
                    \State Obtain $b$ pivots $[\mU^{[J]},\mS^{[J]},\mP^{[J]}] = \aQrcp(\mB^{[J]},b)$.
                    \State Permute $\mA^{[J]} = \mA^{[J-1]}\mP^{[J]}$, $\mW_1^{[J]T}=\mW^{[J-1]T}\mP^{[J]}$, and leading $\mR$ rows.
                    \State \textbf{Construct selected columns $\mAh_J$ from $\mA^{[J]} - \mY^{[J-1]} \mW_1^{[J]T}$.}
                    \State Form reflectors $\mY_2^{[J]}$ using $[\mQ^{[J]},\mR_{11}^{[J]}] = \aQr(\mAh_J)$.
                    \State \textbf{Form inner products $\mW_2^{[J]} = \mT_2^{[J]T} ( \mY_2^{[J]T} \mA^{[J]} - (\mY_2^{[J]T} \mY^{[J-1]}) \mW_1^{[J]T} )$.}
                    \State Augment $\mY^{[J]} = [\mY^{[J-1]} \enspace \mY_2^{[J]}]$ and $\mW^{[J]} = [\mW_1^{[J]} \enspace \mW_2^{[J]}]$.
                    \State \textbf{Construct new rows of $\mR$ from $\mA^{[J]} - \mY^{[J]} \mW^{[J]T}$.}
                    \State Update the sample $\mB_{1}^{[J]}=\mS_{12}^{[J]}-\mS_{11}^{[J]}\mR_{11}^{[J]-1} \mR_{12}^{[J]}$ and $\mB_{2}^{[J]}=\mS_{22}^{[J]}$.
                \EndFor
                \State $\mQ = \mQ^{[1]} \left[ \begin{array}{cc}
                         \mI_{b}&  \\
                         & \mQ^{[2]} \\
                \end{array} \right]  \ldots  \left[ \begin{array}{cc}
                         \mI_{\left([k/b]-1\right)\,b}&  \\
                         & \mQ^{[k/b]}\\
                \end{array} \right]  $.
                \State $\mP = \mP^{[1]} \left[ \begin{array}{cc}
                         \mI_{b}&  \\
                         & \mP^{[2]} \\
                \end{array} \right]  \ldots  \left[ \begin{array}{cc}
                         \mI_{\left([k/b]-1\right)\,b}&  \\
                         & \mP^{[k/b]}\\
                \end{array} \right]$.
        \EndFunction
        \end{algorithmic}
\end{algorithm}

\subsection{Truncated SVD Approximation}

TRQRCP naturally extends to an approximation of the truncated SVD by following the QLP method proposed by Stewart.
The QLP decomposition proceeds by first applying QRCP to obtain $\mA \mP_0 = \mQ_0 \mR$.
Then the right triangular matrix $\mR$ is factored again using an LQ factorization $\mP_1 \mR = \mL \mQ_1$, where row-pivoting is an optional safeguard (otherwise $\mP_1=\mI$),
giving the factorization $\mA=(\mQ_0 \mP_1^T) \mL (\mQ_1 \mP_0^T)$.
The diagonal elements of $\mL$ approximate the singular values of $\mA$.
Huckaby and Chan~\cite{Huckaby2003} provide convergence analysis.

The approximate truncated SVD we propose (TUXV) simply adapts low-rank versions of the steps in QLP.
The rank-$k$ approximation that results is exactly the same as the truncated approximation that would be obtained if QLP had been processed to completion using RQRCP, without secondary row-pivoting, and then truncated to a rank-$k$ approximation.

We begin by using TRQRCP to produce $k$ left reflectors, which defines the initial left orthogonal matrix $\mU^{(0)}$.
Superscript $(0)$ refers to the initial state of an array upon entry to the first iteration of the main loop.
We can compare our results to what would have been obtained from full RQRCP-based QLP:
\begin{align*}
& \mA \mP^{(0)} \approx
   \begin{bmatrix} \mU^{(0)}_1 & \mU^{(0)}_2 \end{bmatrix}
   \begin{bmatrix} \mR^{(0)}_{11} & \mR^{(0)}_{12} \\ 0 & 0 \\ \end{bmatrix} \;\; 
\text{vs.}\;\;
& \hspace{-0.1in}\mA \mP^{(0*)} =
   \begin{bmatrix} \mU^{(0)}_1 & \mU^{(0*)}_2 \end{bmatrix}
   \begin{bmatrix} \mR^{(0)}_{11} & \mR^{(0*)}_{12} \\ 0 & \mR^{(0*)}_{22} \\ \end{bmatrix}.
\end{align*}
The asterisk denotes additional pivoting produced from the full factorization.
The first $k$ pivots in $\mP^{(0)}$ and corresponding reflectors in $\mU^{(0)}$ are the same,
as are corresponding rows in $\mR^{(0)}$ modulo additional column permutations.
We reverse these permutations to construct the $k \times n$ matrix $\mZ^{(0)}=\mR^{(0)} \mP^{(0)T}$ so that
\begin{align*}
\mA \approx
   \begin{bmatrix} \mU^{(0)}_1 & \mU^{(0)}_2 \end{bmatrix}
   \begin{bmatrix} \mZ^{(0)}_{11} & \mZ^{(0)}_{12} \\ 0 & 0 \\ \end{bmatrix}
\quad\text{versus}\quad
\mA =
   \begin{bmatrix} \mU^{(0)}_1 & \mU^{(0*)}_2 \end{bmatrix}
   \begin{bmatrix} \mZ^{(0)}_{11} & \mZ^{(0)}_{12} \\ \mZ^{(0*)}_{21} & \mZ^{(0*)}_{22} \\ \end{bmatrix}.
\end{align*}
Taking the LQ factorization from $\mZ^{(0)}$, so that $\mL^{(1)} \mV^{(1)T} = \mZ^{(0)}$, instead of from $\mR^{(0)}$ simply absorbs the permutation $\mP^{(0)T}$ into the definition of $\mV^{(1)T}$
so that
\begin{align*}
& \mA \approx
   \begin{bmatrix} \mU^{(0)}_1 & \mU^{(0)}_2 \end{bmatrix}
   \begin{bmatrix}
   \mL^{(1)}_{11} & 0 \\
   0 & 0 \\
   \end{bmatrix}
   \begin{bmatrix}
   \mV^{(1)T}_1 \\
   \mV^{(1)T}_2 \\
   \end{bmatrix} \\
\text{versus}\quad
& \mA =
   \begin{bmatrix} \mU^{(0)}_1 & \mU^{(0*)}_2 \end{bmatrix}
   \begin{bmatrix}
   \mL^{(1)}_{11}  & 0 \\
   \mL^{(1*)}_{21} & \mL^{(1*)}_{22} \\
   \end{bmatrix}
   \begin{bmatrix}
   \mV^{(1)T}_1 \\
   \mV^{(1*)T}_2 \\
   \end{bmatrix}.
\end{align*}
For consistency with the following algorithm, we label the $k \times k$ connecting matrix $\mL^{(1)}_{11}$ as $\mX^{(0)}$.
We will discuss the connecting matrix further after explaining the rest of the algorithm.
Provided that no secondary row-pivoting is considered,
the leading $k$ reflectors in $\mV^{(1)}$ and $\mV^{(1*)}$ are identical because they are only computed from the leading $k$ rows of $\mZ^{(0)}$.
At this point, the rank-$k$ approximation of RQRCP-based QLP would require $\mL^{(1*)}_{21}$, which is unknown.
Fortunately, the leading $k$ columns of $\mU^{(0*)} \mL^{(1*)}$ can be reconstructed with one matrix multiply.
We label this $m \times k$ matrix $\mZ^{(1)}$, which is
\[
        \mZ^{(1)} = \mA \mV^{(1)}_1 =
        \begin{bmatrix} \mU^{(0)}_1 & \mU^{(0*)}_2 \end{bmatrix}
    \begin{bmatrix}
      \mL^{(1)}_{11}  \\
      \mL^{(1*)}_{21} \\
    \end{bmatrix}
\]
in both cases.
We can then take the QR-factorization $\mU^{(1)} \mX^{(1)} = \mZ^{(1)}$ to produce the approximation
\begin{equation}
        \mA \approx
                \mU^{(1)}
        \begin{bmatrix}
        \mX^{(1)} & 0 \\
      0       & 0 \\
    \end{bmatrix}
    \mV^{(1)^T}.
    \label{eq:TUXV}
\end{equation}

Further iterations can be computed to produce subsequent $k \times k$ connection matrices $\mX^{(2)}$, $\mX^{(3)}$, etc.~which would flip-flop between upper triangular and lower triangular forms.
To do this, we simply multiply the leading rows of $\mU^T$ or columns of $\mV$ on the left and right of $\mA$, respectively,
as outlined in \cref{alg:TUXV}.
The leading singular values of $\mA$ are approximated on the diagonals of $\mX^{(j)}$.
Since the connection matrix is small, however, it is feasible to obtain slightly better approximations by taking the SVD of $\mX^{(j)}$.
One could also insert mechanisms to iterate until a desired level of convergence is obtained, but Stewart observed that only one QRCP-LQ iteration is needed to produce a reasonable approximation of the SVD.
One subtle point of possible confusion is that by setting $j_{\text{\scriptsize{max}}}=1$ our algorithm might appear to produce a truncated approximation from the sequence RQRCP-LQ-QR.
It is true that the diagonal elements in $\mX$ correspond to that sequence; however, the resulting factorization is equivalent to that which would be obtained by keeping only the leading columns of $\mL$ after RQRCP-LQ.
The final QR factorization simply extracts an orthogonal basis $\mU$.
In the next section, we test the performance of TUXV with $j_{\text{\scriptsize{max}}}=1$ for both timing and quality experiments.

\begin{algorithm}[htb]
        \caption{TUXV approximation of the truncated SVD}
        \label{alg:TUXV}
        \begin{algorithmic}[1]
                \Require
                        \Statex $\mA$ is an $m \times n$ matrix to approximate.
                        \Statex $k$ is the approximation rank.
$k\ll\min(m,n)$.
                        \Statex $j_{\text{\scriptsize{max}}}$ is the number of LQ-QR iterations.
We set $j_{\text{max}}=1$.
                \Ensure
                    \Statex $\mU$ is an orthogonal $m \times m$ matrix.
                    \Statex $\mV$ is an orthogonal $n \times n$ matrix.
                    \Statex $\mX$ is a $k \times k$ upper or lower triangular matrix.
                    \Statex $\mA \approx \mU(:,1:k) \mX \mV(:,1:k)^T$.

                \Function {$[\mU,\mX,\mV]=$ TUXV}{$\mA,k,\tau,j_{\text{\scriptsize{max}}}$}
                    \State \textbf{TRQRCP-Factorize} $[\mU^{(0)},\mR^{(0)},\mP^{(0)}]=\aTrqrcp(\mA,k)$.
                    \State Restore original column order $\mZ^{(0)}=\mR^{(0)}\mP^{(0)T}$.
                        \State \textbf{LQ-Factorize} $[\mV^{(1)},\mX^{(0)T}] = \aQr(\mZ^{(0)T})$.
                    \For{$j=1,3,5,\ldots$}
                        \State $\mZ^{(j)}=\mA \mV^{(j)}(\texttt{:,1:k})$.
                        \State \textbf{QR-Factorize} $[\mU^{(j+1)},\mX^{(j)}] = \aQr(\mZ^{(j)})$.
                        \State If $j=j_{\text{\scriptsize{max}}}$, then break.
                        \State $\mZ^{(j+1)}=\mU^{(j+1)}(\texttt{:,1:k})^T \mA$.
                        \State \textbf{LQ-Factorize} $[\mV^{(j+2)},\mX^{(j+1)T}] = \aQr(\mZ^{(j+1)T})$.
                        \State If $j+1=j_{\text{\scriptsize{max}}}$, then break.
                     \EndFor
                \EndFunction
        \end{algorithmic}
\end{algorithm}

\section{Experiments}
\label{sec:experiments}

Our first Fortran version of RQRCP used simple calls to BLAS and LAPACK subroutines without directly managing workloads among available cores.
Library implementations of BLAS and LAPACK subroutines automatically distribute the computation to available cores using OpenMP.
Although we knew RQRCP should have nearly the same communication complexity as blocked QR, that version did not compete well with library calls to the LAPACK subroutine DGEQRF, the BLAS-3 QR factorization.
We believe this was due to poor automatic memory coordination between large blocked matrix operations.
In order to provide a convincing demonstration of the efficiency of RQRCP, we had to carefully manage workloads using OpenMP within each phase of the main algorithm.
The following experiments show that our RQRCP and TRQRCP subroutines can be written to require substantially less computation time than the optimized QRCP implementation DGEQP3 available through Intel's Math Kernel Library.

\subsection{Full Factorization Time}
\begin{figure}[h]
\centering
\includegraphics[width=0.69\textwidth]{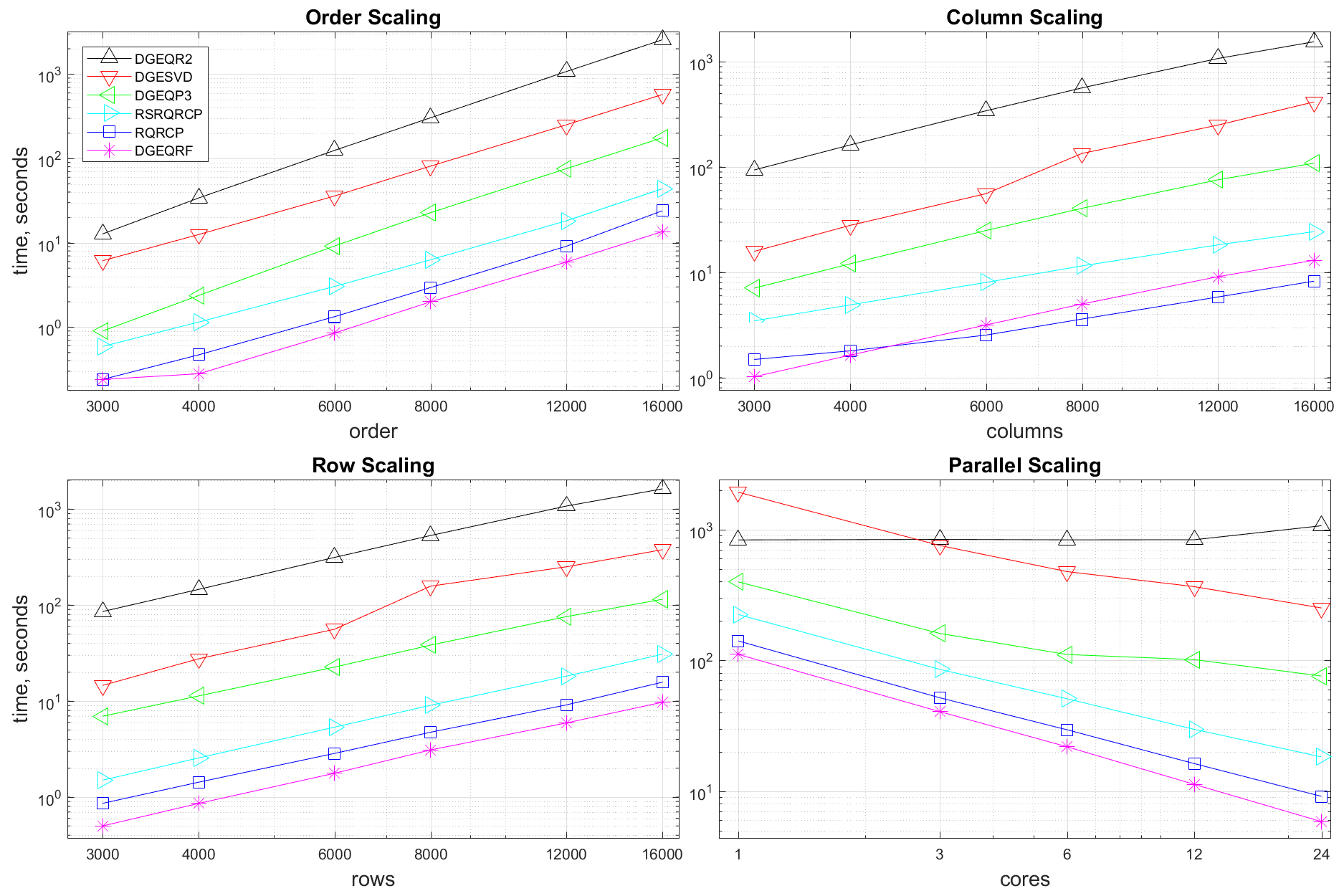}
\caption{Full decomposition benchmarks.
Top-left: $24$ cores, rows and columns scaled equally.
Top-right: $24$ cores, $12000$ rows, columns scaled.
Bottom-left: $24$ cores, rows scaled, $12000$ columns.
Bottom-right: cores scaled, $12000$ rows and columns.
RQRCP consistently performs almost as well as DGEQRF, QR without pivoting.
}
\label{fig:fullTime}
\end{figure}

These tests examine how factorization times scale with various problem dimensions for several full matrix decompositions.
Since we wanted to understand how different sizes and shapes of matrices could affect performance,
we separately tested order scaling, row scaling, and column scaling.
In order scaling, we vary the number of rows and columns together so that the matrix remains square.
We also wanted to see how the performance of each algorithm scales as we increase the number of cores engaged.
Unless the experiment specifies otherwise, each matrix has $12000$ rows, $12000$ columns, and the algorithm engages $24$ cores.
The algorithms we tested are listed in \cref{table1}.
\cref{fig:fullTime} shows performance results.


\begin{table}[h]
\label{table1}
\caption{
Full decomposition experiments compare these algorithms.
Rank-revealing subroutines are DGESVD, DGEQP3, RSRQRCP, and RQRCP.
DGEQRF demonstrates the limit of performance without pivoting.
DGEQR2 to shows the historical evolution of these algorithms.}
\centering
\footnotesize
\setlength{\tabcolsep}{5pt}
\begin{tabular}{|l|l|}\hline
        \text{Subroutine}     & \text{Description} \\ \hline\hline
        \aQrTwo & LAPACK BLAS-2 implementation of QR\\ \hline
        \aSvd & LAPACK singular value decomposition\\ \hline
        \aQrcpThree & LAPACK competing implementation of QRCP\\ \hline
        \aRsrqrcp& \cref{alg:RQRCP} modified for repeated-sampling\\ \hline
        \aRqrcp  & \cref{alg:RQRCP} with sample update (unmodified)\\ \hline
        \aQrThree & LAPACK BLAS-3 implementation of QR\\ \hline
\end{tabular}
\end{table}

These tests show that RQRCP performs nearly as well as DGEQRF, the LAPACK implementation of BLAS-3 QR without pivoting.
We further note that our implementation even outperforms DGEQRF in some column scaling cases. 
These experiments were run on a single node of the NERSC machine Edison.
Each node has two 12-core Intel processors.
Subroutines were linked with Intel's Math Kernel Library.
Each test matrix was randomly generated and the same matrix was submitted to each algorithm.

\subsection{Truncated Decomposition}

\begin{figure}[h]
\centering
\includegraphics[width=0.69\textwidth]{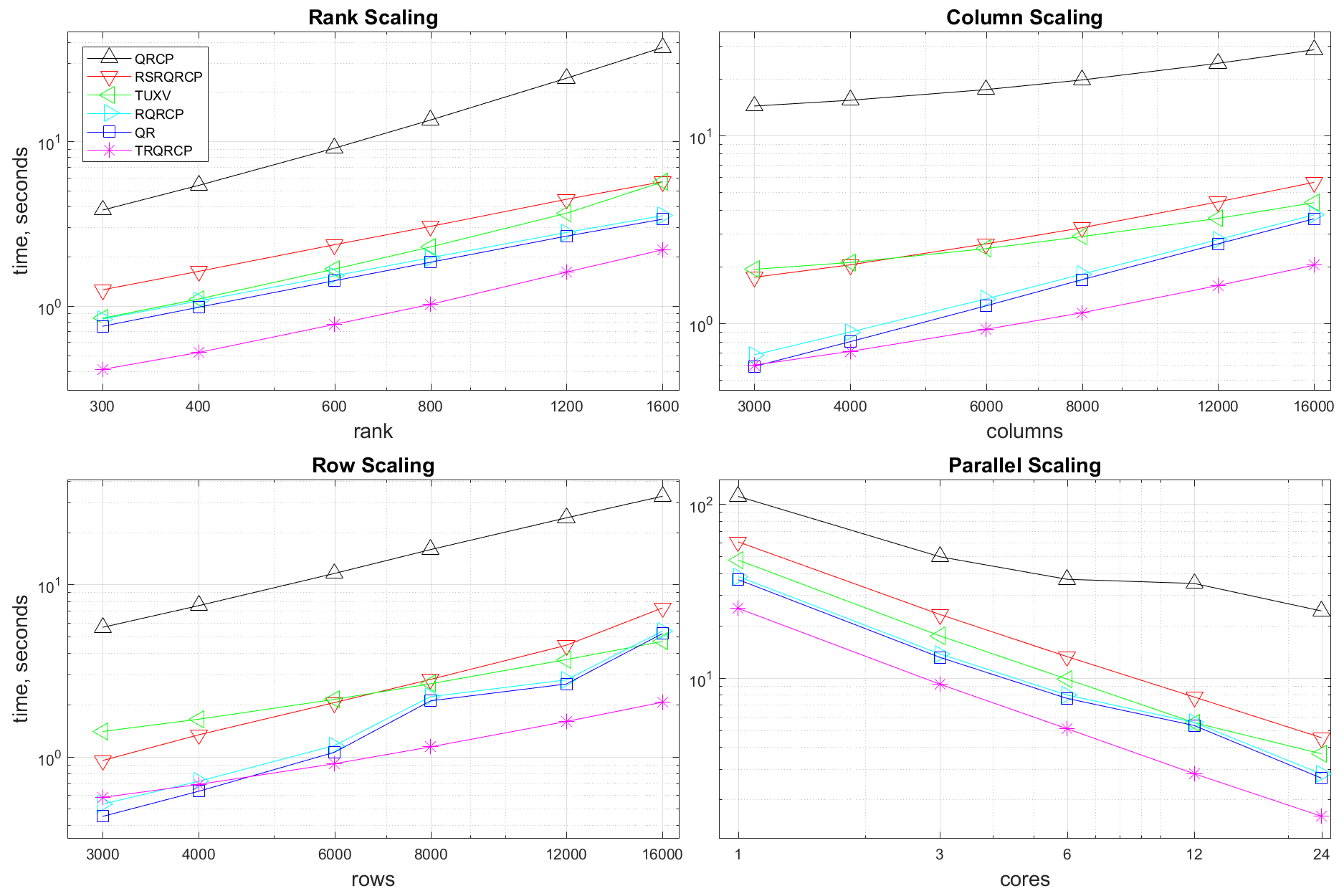}
\caption{Truncated decomposition benchmarks.
Top-left: $24$ cores, $12000$ rows and columns, truncation rank scaled.
Top-right: $24$ cores, $12000$ rows, columns scaled, rank $1200$.
Bottom-left: $24$ cores, rows scaled, $12000$ columns, rank $1200$.
Bottom-right: cores scaled, $12000$ rows and columns, rank $1200$.
RQRCP and QR perform similarly.
TRQRCP is fastest.
TUXV requires only modest additional cost.
}
\label{fig:truncTime}
\vspace{-0.15in}
\end{figure}

These experiments compare truncated approximation performance.
In addition to probing how the matrix shape affects execution time,
we examine the effect of varying the truncation rank.
As before, we test parallelization by varying the number of cores engaged.
Unless specified otherwise, each algorithm engages $24$ cores, operates on a $12000 \times 12000$ matrix, and truncates to rank $1200$.
Since the proprietary optimized implementations of LAPACK functions were unavailable for modification,
we rewrote and adjusted each algorithm to halt at the desired rank.
\Cref{table3} lists the algorithms we tested. 
Again, the same random matrix is submitted to each algorithm.
\Cref{fig:truncTime} shows results.

\begin{table}[h]
\label{table3}
\caption{These algorithms are compared in truncated decomposition scaling experiments.
Comparing RSRQRCP with RQRCP reveals the cost of repeated sampling.
Comparing RQRCP with QR reveals the cost of pivot selection from samples.
Comparing RQRCP with TRQRCP reveals the cost of the trailing matrix update.
This version of QR is identical to RQRCP after eliminating sample operations and pivoting.}
\centering
\footnotesize
\setlength{\tabcolsep}{5pt}
\begin{tabular}{|l|l|}\hline
        \text{Subroutine}     & \text{Description} \\ \hline\hline
        \aQrcp    & \cref{alg:QRCP3}, QRCP with trailing update\\ \hline
        \aRsrqrcp & \cref{alg:RQRCP} modified for repeated-sampling\\ \hline
        \aTuxv    & \cref{alg:TUXV}, approximation of truncated SVD\\ \hline
        \aRqrcp   & \cref{alg:RQRCP} with sample update and trailing update (unmodified)\\ \hline
        \aQr      & QR (no pivoting) with trailing update\\ \hline
        \aTrqrcp  & \cref{alg:TRQRCP} with sample update and no trailing update\\ \hline
\end{tabular}
\end{table}

Since TRQRCP uses the sample update formula and avoids the trailing update, it is nearly always fastest. 
Our TUXV experiments use $j_{\text{\scriptsize{max}}}=1$.
As such, TUXV performs just one additional matrix multiply.
These results show that TUXV requires only a modest increase in processing time over optimized truncated QR.
Furthermore, the next set of experiments shows that TUXV gains a significant improvement in approximation quality over both QRCP and RQRCP.

\subsection{Decomposition quality}
\label{subsec:decompquality}
\begin{figure}[h]
\centering
\includegraphics[width=0.69\textwidth]{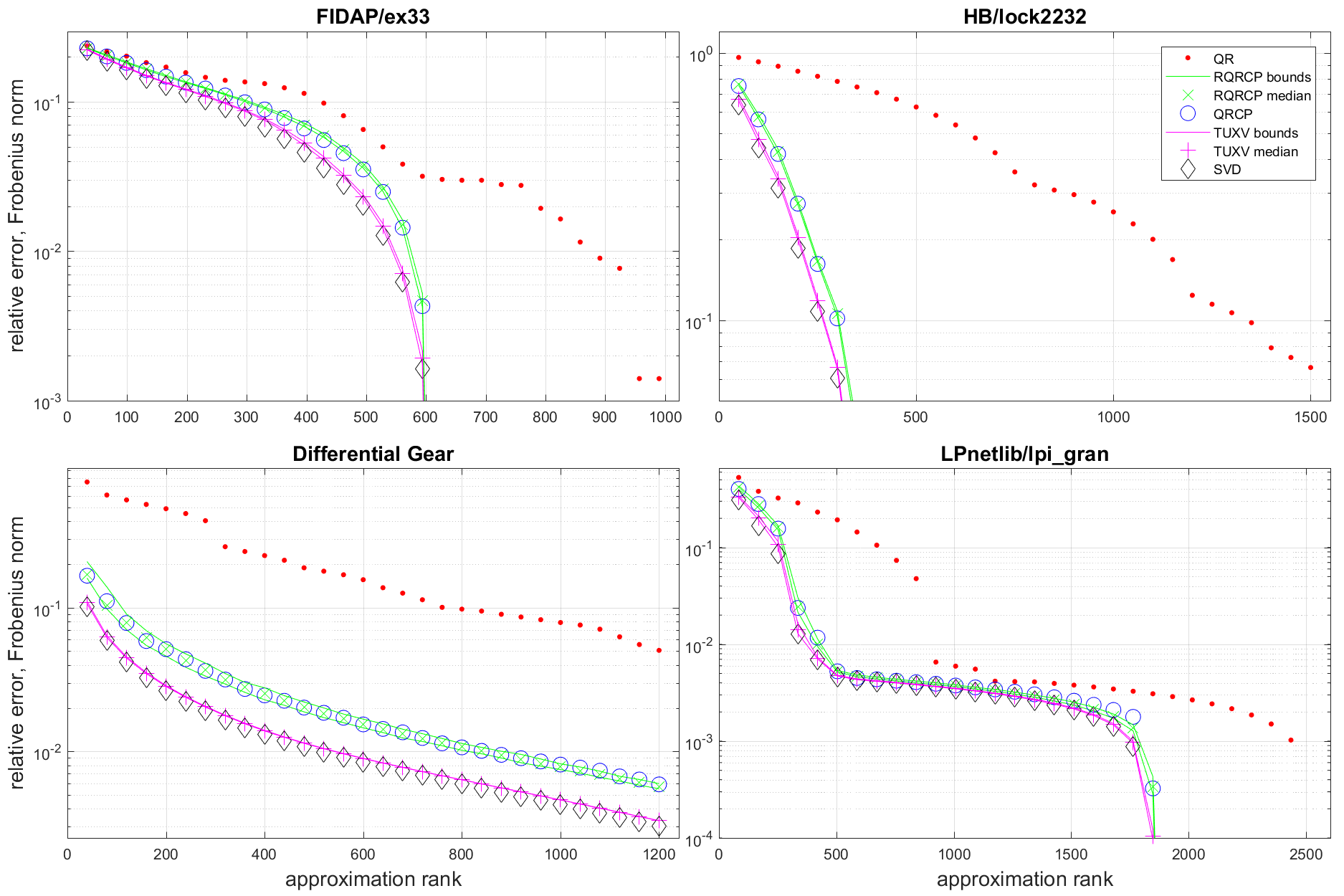}
\caption{Truncated decomposition quality experiments.
Top-left: {\texttt{FIDAP/ex{{33}}}} $(1733 \times 1733)$.
Top-right: {\texttt{HB/lock{{2232}}}} $(2232 \times 2232)$.
Bottom-left: Differential gear $(2442 \times 3888)$.
Bottom-right: {\texttt{LPnetlib/lpi\_gran}} $(2658 \times 2525)$.
Both RQRPC and TUXV show the minimum and maximum truncation errors over 100 runs.
The range of outcomes for RQRCP is narrow and holds to QRCP.
TUXV outcomes are even narrower and slightly above the truncated SVD. 
}
\label{fig:approxError}
\end{figure}

The pivots that result from randomized sampling are not the same as those obtained from QRCP.
In order to compare factorization quality, we construct sequences of partial factorizations and compute the corresponding low rank approximations.
The resulting relative error in the Frobenius norm is plotted in \cref{fig:approxError} against the corresponding approximation rank.
For both RQRCP and TUXV, we perform 100 runs and plot the median, as well as both the minimum and maximum error bounds. 
Random samples used padding size $p=8$ and block size $b=32$. 
Matrix decomposition quality is compared for the proposed algorithms using test cases from the San Jose State University Singular Matrix Database: \texttt{FIDAP/ex33}, \texttt{HB/lock2232}, and \texttt{LPnetlib/lpi\_gran}.
We also test a matrix corresponding to a gray-scale image of a differential gear (image credit: Alex Kovach~\cite{Kovach2016}).
Plot axes have been chosen to magnify the differences among these algorithms.

At the top of each plot we have QR without pivoting.
In order to produce competitive results, QR was applied after presorting columns in order of descending 2-norms.
Despite this modification, QR performs poorly (as expected) with approximation error dropping much more slowly than the other approximations,
thus demonstrating why pivoting is necessary to prioritize representative columns in the truncated decomposition.
Below QR we have RQRCP, which achieves results that are consistent with QRCP in each case.
The QRCP-like low-rank approximations are further improved by TUXV and the exact truncated SVD, which is theoretically optimal.
In each case, TUXV produces approximation error closer to the SVD, with very low variation among the 100 runs.

\begin{figure}[h]
\centering
\includegraphics[width=0.69\textwidth]{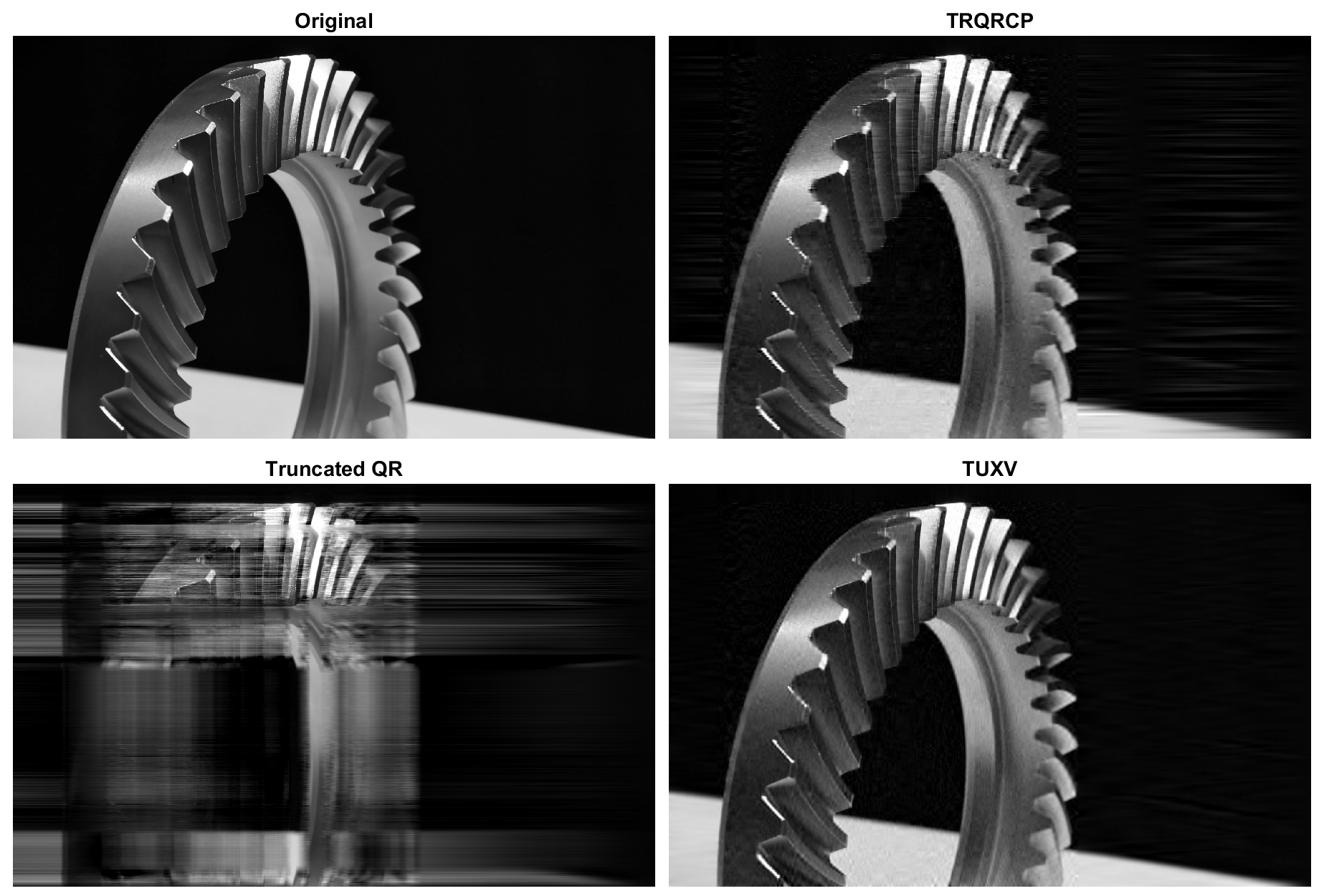}
\caption{
Low-rank image reconstruction comparison.
Reconstructions are computed from a $2442 \times 3888$ grayscale matrix.
Approximations are reconstructed with rank $80$.
Truncated QR is computed with columns presorted by descending $2$-norm.
TRQRCP dramatically improves visual approximation quality.
Likewise, TUXV further improves fine details.
}
\label{fig:gear}
\end{figure}

In \cref{fig:gear}, we compare approximation quality by reconstructing the image of the differential gear using low-rank approximations.
The original image is $2442 \times 3888$ and we truncate to rank $80$.
Again, truncated QR shows the poorest reconstruction quality, despite presorting.
Truncated RQRCP produces better results, but close inspection shows fine defects.
Reconstruction using TUXV is nearly indistinguishable from the original.

\section{Conclusion}
\label{sec:conclusion}

We have shown that RQRCP achieves strong parallel scalability and the pivoting quality of QRCP at BLAS-3 performance,
often an order of magnitude faster than the standard approach.
By using randomized projection to construct a small sample matrix $\mB$ from a much larger original matrix $\mA$,
it becomes possible to substantially reduce the communication complexity associated with a series of algorithmic decisions.
Our analysis of latent column norms, inferred from the sample, justifies the selection computations that we use to obtain full blocks of column pivots.
Having blocks of pivots allows our algorithm to factorize $\mA$ using matrix-matrix multiplications instead of interleaving multiple series of matrix-vector operations,
making RQRCP the algorithm of choice for numerical rank determination.

Critically, we have shown how to leverage intermediate block transformations of $\mA$ to update $\mB$ with a sample update formula.
This technique allows us to avoid computing a new randomized projection for each block operation, 
thus substantially reducing the matrix-matrix multiplication work needed to process each block.

We have extended this method of factorization to produce truncated low-rank approximations.
TRQRCP, the truncated formulation of RQRCP, avoids block updates to the trailing matrix during factorization,
which reduces the leading contribution to communication for low-rank approximations.
Moreover, TRQRCP provides an efficient initial operation in our approximation of the truncated SVD, TUXV.

Our algorithms are implemented in Fortran with OpenMP.
Numerical experiments compare performance with LAPACK subroutines, linked with the Intel Math Kernel Library, using a 24-core system.
These experiments demonstrate that the computation time of RQRCP is nearly as short as that of unpivoted QR and substantially shorter than QRCP.
Problems that have been too large to process with QRCP-dependent subroutines may now become feasible.
Other applications that had to settle for QR, due to performance constraints, may find improved numerical stability at little cost by switching to RQRCP.
For low-rank approximations, TRQRCP offers an additional performance advantage and TUXV improves approximation quality at a modest additional cost.
These algorithms open a new performance domain for large matrix factorizations that we believe will be useful in science, engineering, and data analysis.

Randomized methods harness the fact that it is possible to make good algorithmic decisions in the face of uncertainty.
By permitting controlled uncertainty, we can substantially improve algorithm performance.
Future work will address how we may understand the foundations of credible uncertainty in predictions.
Improving this understanding will facilitate the development of efficient predictive algorithms and support our ability to make good decisions from limited information.

\section*{Acknowledgments}
We sincerely thank Chris Melgaard, Laura Grigori, and James Demmel for several helpful discussions, and to anonymous reviewers for their hard work and insightful feedback.
We also deeply appreciate feedback on this revision from both Thomas Catanach and Jacquilyn Weeks.

Sandia National Laboratories is a multimission laboratory managed and operated by National Technology and Engineering Solutions of Sandia, LLC., a wholly owned subsidiary of Honeywell International, Inc., for the U.S. Department of Energy's National Nuclear Security Administration under contract DE-NA-0003525.

This paper describes objective technical results and analysis. Any subjective views or opinions that might be expressed in the paper do not necessarily represent the views of the U.S. Department of Energy or the United States Government.

\bibliographystyle{siamplain}
\bibliography{references}

\end{document}